\documentclass[aps,prl,showpacs,amssymb,amsmath]{revtex4}
\usepackage[dvips]{graphicx,color}
\usepackage{color}
\begin{document}
\title{Shot noise in charge and magnetization currents of a quantum ring.}
\author{Fabio Cavaliere$^{1,2}$, Federica Haupt$^{2}$, Rosario
  Fazio$^{3}$, and Maura Sassetti$^{2}$ \vspace{1mm}} 
\affiliation{
  $^{1}$ I. Institut f\"ur Theoretische Physik, Universit\"at Hamburg,
  Jungiusstra\ss{}e 9, 20355 Hamburg, Germany\\
  $^{2}$ LAMIA-INFM and Dipartimento di Fisica, Via Dodecaneso 33, 16146 Genova, Italy \\
  $^{3}$ NEST-INFM and Scuola Normale Superiore, Piazza dei Cavalieri
  7, I-56126 Pisa, Italy\\ \vspace{3mm}} \date{December 10, 2004}
\begin{abstract} 
  The shot noise in a quantum ring, connected to leads, is studied in
  the presence of electron interactions in the sequential tunneling
  regime. Two qualitatively different noise correlations with
  distinctly different behaviors are identified and studied in a large
  range of parameters. Noise in the total current is due to the
  discreteness of the electron charge and can become super-Poissonian
  as result of electron interaction. The noise in the
  magnetization current is comparatively insensitive to the
  interaction but can be greatly enhanced if population inversion of
  the angular states is assumed. The characteristic time scales are
  studied by a Monte-Carlo simulation.
\end{abstract}
\pacs{73.50.Td,71.10.Pm,73.23.-b}
\maketitle
%%%%%%%%%%%%%%%%%%%%%%%%%%%%%%%%%%%%%%%%%%%%%%%%%%%%%%%%%%%%%%%%%%%%%%%%
 \noindent 
 Noise is known to be a key tool to study non-equilibrium transport
 properties of mesoscopic quantum systems~\cite{buttiker}.  In the
 regime of Coulomb Blockade the presence of electron interaction
 generally leads to a {\em suppression} of the noise as compared with
 the Poissonian limit~\cite{herkor}. Recently, several mechanisms have
 been proposed leading to an {\em enhancement} of the shot noise.  A
 super-Poissonian regime has been found in the presence of a negative
 differential conductance~\cite{kiesslich}, for resonant
 tunneling~\cite{haug} and via localized states~\cite{safonov}.
 Super-Poissonian noise is also expected in a few-level quantum dot in
 the presence of a magnetic field~\cite{belzig}.

In this work, we study current correlations in a one-dimensional (1D)
quantum ring, connected via tunnel contacts to external leads, in the
presence of interactions. The system is characterized by discrete
charge and angular momentum degrees of freedom, and  can be used to 
study interaction-induced noise effects of an unprecedented richness, 
in the presence of an Aharonov-Bohm flux threading the ring. 
We characterize the noise by considering both the {\em charge 
current noise} (related to the fluctuations of the 
tunneling current) and the {\em magnetization current noise} (related
to the fluctuations in the persistent current). 
Our main results are: (i) regions of super-Poissonian charge 
noise can be found whenever high orbital channels contribute to transport. 
(ii) Charge noise is sensitive to asymmetry of the
 tunnel barriers; by changing the Aharonov-Bohm flux the charge noise
 can be tuned from sub- to super-Poissonian.
(iii) The magnetization noise is considerably stable against
 asymmetry and the presence of interactions.  In order to enter a regime
 of high magnetization noise, it is necessary to have {\em population
   inversion} in the orbital states.  

All of the phenomena discussed
below {\em cannot} be observed in a non-interacting system and should
 be accessible with presently available experimental technology.
 
 For the isolated quantum ring we assume a spinless
 Luttinger liquid (LL) with density-density electron interactions ($g$
 interaction parameter, $g=1$ no interaction)~\cite{voit}
\begin{equation}
\label{eq:1}
H_{\rm ring}=\frac{E_{N}}{2}
{\left({N}-N_{\rm g}\right)}^2+\frac{E_{J}}{2} 
{\left(J-2\frac{\Phi}{\Phi_{0}}\right)}^2
+\sum_{n>0}n \varepsilon\,b^{\dagger}_{n} b_{n}\,. 
\end{equation}
The zero mode ${N}={N_+}+{N_-}$ represents the total number excess of
electrons, ${J}={N_+}-{N_-}$ gives the imbalance between clockwise
($N_+$) and anticlockwise ($N_-$) moving electrons and $N$, $J$ have
integer eigenvalues with selection rules $(-1)^N=(-1)^J$ ~\cite{loss}.
The gate charge is $N_{\rm g}$ and the flux is $\Phi$, with
$\Phi_0=h/e$ the flux quantum. The charging energy $E_{N}$ is
mainly due to the Coulomb interaction and influenced by the coupling
with the external circuit.  It is the largest energy scale in the
model. The orbital addition energy is $E_{J}=\pi v_{\rm F}/L\ll E_{N}$
($v_{\rm F}$ Fermi velocity, $L$ circumference of the ring).  The last
term in Eq.(\ref{eq:1}) represents plasmons moving with a renormalized
velocity $v= v_{\rm F}/g$ and excitation energy $\varepsilon=2\pi
v/L$.

The ring is connected to external leads via tunneling barriers. This
creates charge, $I_{N}^{(i)}$, and orbital, $I_{J}^{(i)}$, tunnel
currents related to the clockwise and anticlockwise contributions:
$I_{N}^{(i)}=I^{(i)}_{+}+I^{(i)}_{-}$ and
$I_{J}^{(i)}=I^{(i)}_{+}-I^{(i)}_{-}$ at junctions ($i=1,2$).  The
tunneling processes modify charge and magnetization via
$\dot{N}\propto I_{N}$ and $\dot{J}\propto I_{J}$.

We evaluate the zero frequency correlators ($\nu = {N,J}$)
\begin{equation}
S^{(ij)}_{\nu}= \int_{-\infty}^{\infty}{\rm d t}\ \langle\Delta
I^{(i)}_{\nu}(t) \Delta I^{(j)}_{\nu}(0)+ \Delta I^{(i)}_{\nu}(0)
\Delta I^{(j)}_{\nu}(t)\rangle
\end{equation}
at terminals $i$ and $j$, between current fluctuations $\Delta
I^{(i)}_{\nu}(t)=I^{(i)}_{\nu}(t)-\langle I^{(i)}_{\nu}\rangle$.  The
knowledge of $S_N$ and $S_J$ are ingredients for determining
the auto- ($S_{++}$, $S_{--}$) and cross-correlators ($S_{+-}$,
$S_{-+}$).

The leads are assumed as interacting LLs with interaction parameter
$g_{\ell}$.  They can be realized as edge states of fractional
quantum Hall liquids~\cite{wen} or in quantum wires~\cite{yacoby}. The
difference of their chemical potential is controlled by the external
bias voltage $\mu_2-\mu_1= eV$.

\begin{figure}[htpb]
  \setlength{\unitlength}{1cm}
  \includegraphics[clip,width=7.6cm,keepaspectratio]{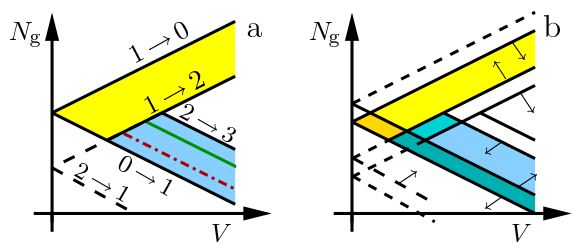}
  \vskip-0.1cm
\caption[]{Scheme of conducting regions in the $(V,N_{\rm g})$
  plane. (a) $\Phi=0$. Black lines: onset of the
  transitions $|J|\to |J'|$. Solid lines: detectable in 
  current. Yellow: $J_{\rm max}=1$; cyan:
  $J_{\rm max}=2$. (b) $0<\Phi<\Phi_0/2$. Each line of (a) is now split in two.
  Arrows indicate lines shift with increasing
  $\Phi$.}
\label{fig1}
\end{figure}

For weak coupling (level broadening due to tunneling much smaller than
temperature and level spacing) the electronic states  $\left|\alpha\right>$ are
occupied with probability $P_{\alpha}(t)$. We 
assume fast plasmon relaxation (see Ref.\cite{kim} for the opposite 
situation) towards thermal equilibrium, and we consider $k_{\rm B}T,eV\ll
E_{N}$ so that the transport is governed just by two charge states
$N$ and $N+1$. The ring states are then completely characterized by their 
angular degree of freedom $J$. From now on, the index $N$ is omitted. 
In the non-linear regime ($eV>E_{J}$) many 
orbital states $J_{\rm min}\leq J\leq J_{\rm max}$ contribute to  
transport. For $\Phi=0$, $J_{\rm max}=-J_{\rm min}$. Schemes
of the relevant transport states are shown in Fig.~\ref{fig1}.

The vector ${\bf P}\equiv\{P_{J_{\rm min}},\ldots, P_{J_{\rm max}}\}$
represents the occupation probabilities. It is governed by the master
equation $\dot{\bf P}(t)=M\cdot{\bf P}(t)$. The square matrix $M$
contains the tunneling rates $\gamma^{(i)}_{J\to J'}$~\cite{europhys}
through the $i$-th barrier with $J'=J\pm 1$.  Following
~\cite{herkor}, we calculate the correlation functions $S_{\nu}^{(ij)}$
by spectrally decomposing $M$.  Because of charge conservation 
$\langle I_{\nu}^{(i)}\rangle=\langle I_{\nu}\rangle$, and 
$S_{\nu}^{(ij)}=S_{\nu}$ independently of the barrier indices.

We first consider $\Phi=0$ and voltages that allow $J_{\rm max}=2$
(Fig.~\ref{fig1}a).  We consider the charge current Fano factor
$F_{N}=S_{N}/2e|\!\left<I_{N}\right>\!|$ and define an angular current
Fano factor $F_{J}=S_{J}/2e|\!\left<I_{N}\right>\!|$ related to
fluctuations of $I_{J}$. We quote, for the discussion, the zero
temperature case which, however, describes very accurately the
temperature regime $k_{\rm B}T<0.1 E_{J}$
\begin{eqnarray}
\label{fano}
\!\!F_{N}\!&=&\!1+2\ \frac{r_1(1-r_2)^2-(1+r_1r_2)(1+r_1)\tau_{1}/\tau_{0}}
{\left[1+r_1r_2+(1+r_1)\tau_{1}/\tau_{0}\right]^2}\\\nonumber\\
\!\!F_{J}&=&1-2r_1(1-r_1)/(1+r_1)\label{fanoj}\\\nonumber\\
\!\!r_1&=&\gamma_{1\to2}^{(1)}/\gamma_{1\to0}^{(1)}\,,
\quad r_2=2\gamma_{0\to1}^{(2)}/
\gamma_{2\to1}^{(2)}=\tau_2/\tau_0\,.
\label{ratio}
\end{eqnarray}
Here, $r_2$ represents the escape ratio from $J=0$ and $|J|=2$, $r_1$
is the population ratio between $|J|=2$ and $J=0$, and $r_1\cdot r_2$
is the stationary occupation probability ratio $2P_2/P_0$. The
corresponding dwell times $\tau_{|J|}$ are:
$\tau_{0}^{-1}=2\gamma_{0\to 1}^{(2)}$, $\tau_{1}^{-1}=\gamma_{1\to
  2}^{(1)}+\gamma_{1\to 0}^{(1)}$ and $\tau_{2}^{-1}=\gamma_{2\to
  1}^{(2)}$. For $J_{\rm max}=1$ (Fig.~\ref{fig1}a yellow),  $r_1=0$
and the above expressions reduce to the results for a three channel
system, $F_{N}<1$, and $F_{J}=1$~\cite{herkor,ale,feinberg}, which
implies $F_{+-}\propto F_{N}-F_{J}<0$. For $J_{\rm max}=2$ (Fig.~\ref{fig1}a cyan), $r_1\neq
0$ and a new dynamics is found.
Here, when the degeneracy of the dwell times
$\tau_0$ and $\tau_2$ is lifted, $r_2 \neq 1$, the charge noise becomes 
super-Poissonian, $F_{N}>1$,  for asymmetries of the two barriers larger 
than 
\begin{equation}
\label{eq:critical}
A_{\rm c}=\frac{2\bar{\gamma}_{0\to1}^{(2)}}
{\bar{\gamma}_{1\to0}^{(1)}+\bar{\gamma}_{1\to2}^{(1)}}
\frac{(1+r_1r_2)(1+r_1)}{r_1(1-r_2)^2} 
\end{equation}
(the asymmetry $A$ is defined as the ratio of the high temperature
tunneling resistances of the two barriers). The $\bar{\gamma}$ rates
are defined modulo the latter resistances.  The angular Fano factor
$F_{J}$ depends {\em only} on $r_1$ and $F_{J} >1$ can be achieved
when $r_1>1$, i.e. with a {\em population inversion} between transitions $1\to 2$ and $1\to 0$ (cf. Eq.~(\ref{ratio})).

For interaction $0.5<g<1$, the region with $J_{\rm max}=2$ is split by the line $E_{2\to1}=e V/2+E_{N} (N_{\rm g}-1/2)+ 3
E_{J}/2=\varepsilon$ (green line in Fig.~\ref{fig1}a), corresponding
to the transition $2\to1$ with a plasmonic excitation in the final
state. We denote these regions as $I$ and $II$, depending on
$E_{2\to1}<\varepsilon$ or $>\varepsilon$. If $g=1$, only region $II$ is 
present, if $g<0.5$, there is only region $I$.  

\begin{figure}[htpb]
\setlength{\unitlength}{1cm}
\includegraphics[clip,width=7.6cm,keepaspectratio]{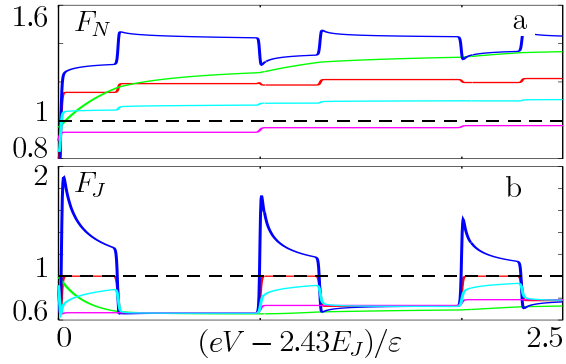}
\vskip-0.1cm 
\caption[]{Fano factors as a function of $(eV-2.43E_J)/{\varepsilon}$, 
  where $V$ varies according to $N_{\rm
    g}=1/2+0.93E_{J}/E_{N}-eV/2E_{N}$, with $k_{\rm B}T=0.02 E_{J}$ and
  $A=20$. (a) $F_{N}$ for an interacting ring, $g=0.7$, and different
  interactions in the leads: $g_{\ell}=1$ (red), 0.9 (cyan), 0.5
  (green), 1.2 (blue); magenta: $g=g_{\ell}=1$. (b) $F_{J}$,
  parameters and colors as in (a).}
\label{fig2}

\end{figure}

Figure~\ref{fig2}
shows charge and magnetization Fano factors along the dashed-dotted
red line in Fig.~\ref{fig1}a, i.e.  as a function of
$(eV-2.43E_J)/{\varepsilon}$, where $V$ varies according to $N_{\rm
  g}=1/2+0.93E_{J}/E_{N}-eV/2E_{N}$.  Without interaction (magenta)
the degeneracy of the plasmon energy, $\varepsilon=2 E_{J}$, implies
$r_2=1$ and $r_1<1$, leading {\em always} to sub-Poissonian noise.
With interaction one has $r_2\neq 1$ so that $F_{N}>1$ if $A>A_{\rm
  c}$. This result is robust against changes of strength and sign of
the interaction. For $F_{J}$, the sign of the interaction is
crucial. Only for attractive leads, $g_{\ell}>1$, it is possible to achieve an inversion of population with $F_{J}>1$
(Fig.~\ref{fig2}b blue).

%In region $I$, for $g<1$ and $g_{\ell}=1$ one always has $r_2=2$ and 
%two energy intervals: $\sigma_-(n)$ given by
%$(n-1)\varepsilon<eV-2.43E_J<(n-g)\varepsilon$ and $\sigma_+(n)$ given by
%$(n-g)\varepsilon<eV-2.43E_J<n\varepsilon$, with $n\ge 1$.  These intervals
%correspond to different population ratios of $r_1$: $r_1=1$ in $\sigma_{-}(n)$
%and $r_1=n/(n+\lambda)$ in $\sigma_{+}(n)$.
In region $I$, for $g<1$ and $g_{\ell}=1$, it is always $r_2=2$ and
one can identify two energy intervals, $\sigma_{\pm}(n)$,
corresponding to different population regions: $r_1=1$ in
$\sigma_{-}(n)$ and $r_1=n/(n+1)$ in $\sigma_{+}(n)$. Referring to
Fig.~\ref{fig2}, $\sigma_{-}(n)$ is given by
$(n-1)\varepsilon<eV-2.43E_{J}<(n-g)\varepsilon$ and $\sigma_{+}(n)$
by $(n-g)\varepsilon<eV-2.43E_{J}<n\varepsilon$. In the first
interval, for $A\gg A_{\rm c}$ the system belongs to a universality
class with values $F_{N}=1+2/9$ and $F_{J}=1$. By tuning the voltage
along the diagonal line one enters periodically the interval
$\sigma_{+}(n)$ where the Fano factors depend on the interaction,
$\lambda=(g+g^{-1})/2$
\begin{equation}
F_{N}=1+\frac{2n(n+\lambda)}{(3n+\lambda)^{2}}\,,\quad
F_{J}=1-\frac{2n\lambda}{(n+\lambda)(2n+\lambda)}\, .
\end{equation}
Interactions in the leads ($g_{\ell}\neq 1)$ modify the above modulation introducing 
a power law behavior at the thresholds of each region. For strong repulsive interaction, $g_{\ell}<0.5$, the
memory of plasmons is completely lost, with an increase
of $F_{N}$ and a depletion of $F_{J}$ (green curves).

\begin{figure}[htpb]
\setlength{\unitlength}{1cm}
\includegraphics[clip,width=7.6cm,keepaspectratio]{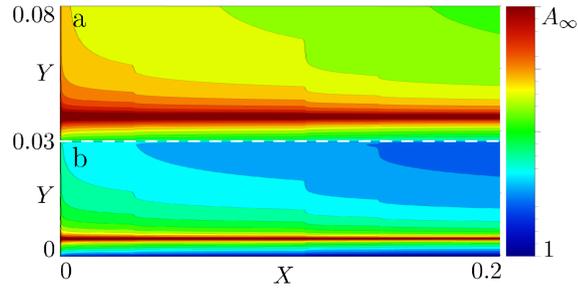}
\vskip-0.1cm
\caption[]{Color-contour plot of $A_{\rm c}$ for $g=0.7$, $g_{\ell}=0.8$,
  $k_{\rm B}T=0.02 E_{J}$ in the $XY$ plane (see text). (a) Region $I$,
  $A_{\infty}=4100$. (b) Region $II$, $A_{\infty}=2.6\ 10^5$.}
\label{fig3}
\end{figure}

In Fig.~\ref{fig3}, the critical asymmetry $A_{\rm c}$ is plotted in
the plane $XY$, where $X=1/2-N_{\rm g}-1.5E_{J}/E_{N}+eV/2E_{N}$ and
$Y=-1/2 -N_{\rm g}-0.5E_{J}/E_{N}+eV/2E_{N}$. Note that $Y=0$
corresponds to the transition line $0\to 1$, $Y=0.03$ to the line
$E_{2\to 1}=\varepsilon$, and $X=0$ to the line $1\to 2$
(cf. Fig.~\ref{fig1}a). In the panels a and b, the regions $I$ and
$II$ are shown.

In both regions, near certain lines parallel to the $X$ axis, we have
$r_2=1$.  Here, $A_{\rm c}$ diverges and no super-Poissonian charge
noise can be achieved.  Away from these lines, increasing the voltage,
$A_{\rm c}$ decreases because of the {\em increasing number of excited
  plasmons} present in the output transition. These randomize the
system and decrease the dwell time $\tau_1$ renormalizing the output
barrier having even $A_{\rm c}<1$ for sufficiently high voltage.

For $\Phi\neq 0$, the degeneracy of the states with $\pm J$ is lifted.
As a consequence, each transition line at $\Phi=0$ splits in two, which move
in opposite directions when increasing $\Phi$
(Fig.~\ref{fig1}b).  This causes several effects in transport regions
where $J_{\rm max}\neq-J_{\rm min}$.

\begin{figure}[htpb]
\setlength{\unitlength}{1cm}
\includegraphics[clip,width=8.3cm,keepaspectratio]{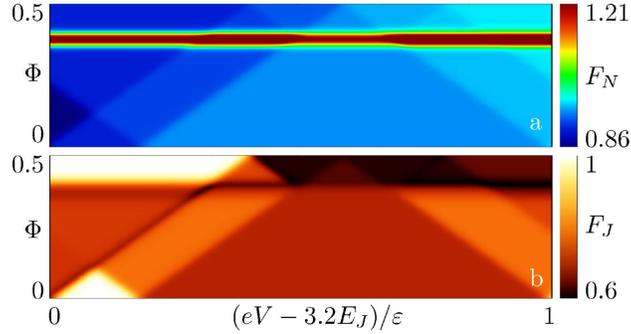}
\vskip-0.1cm 
\caption[]{Density plot of $F_{N}$ (a) and $F_{J}$ (b) as a function of $(eV-3.2E_J)/\varepsilon$
  (voltage moving according to $N_{\rm
    g}=1/2+1.7E_{J}/E_{N}-eV/2E_{N}$) and $\Phi$, in units $\Phi_{0}$.
  Parameters are: $g=0.8$, $g_{\ell}=1$, $k_{\rm B}T=0.02 E_{J}$ and
  $A=20$.}
\label{fig4}
\end{figure}

Figure~\ref{fig4} shows results for $F_{N}$ and $F_{J}$ along a
diagonal line (inside region $II$ at $\Phi=0$) for
$0<\Phi<\Phi_{0}/2$.  Results for $\Phi_{0}/2<\Phi<\Phi_{0}$ are
specular with respect to the $\Phi=\Phi_0/2$ axis and periodic in the
flux with period $\Phi=\Phi_0$.  Increasing $\Phi$, many transitions
cross the fixed line, so that one can study the correlations in a wide
range of transport regions.  As shown in Fig.~\ref{fig4}a, the noise
can change dramatically by tuning the flux.  At $\Phi=0$ one has
$A<A_{\rm c}$, so that $F_{N}<1$. However, increasing the flux at
fixed energy, super-Poissonian noise is reached for $\Phi\approx
0.4\Phi_0$. The onset of this region is at  $\Phi^*=
0.35\Phi_0$, given by the intersection of
the line along where $V$ is varied, $N_{\rm g}=1/2+1.7E_{J}/E_{N}-eV/2E_{N}$, 
with the upper moving transition line $2\to 1$ with one plasmon, $N_{\rm
  g}=1/2+E_{J}/2E_{N}(4/g-3+4\Phi/\Phi_0)-eV/2E_{N}$, i.e. 
$\Phi^*=[1.7+(3-4/g)/2]\Phi_0/2$. 
Inducing transitions between sub- and
super-Poissonian behavior as a function of the flux is strictly a signature of
interactions.  In a noninteracting ring we {\em
  always} find sub-Poissonian behavior, regardless of the number of
states supporting the transport. For the magnetization
current, the interaction in the ring and finite flux are not enough to
induce $F_{J}>1$ (Fig.~\ref{fig4}b).  The flux allows to control
systematically the sign of cross-correlators $F_{+-}\propto
F_{N}-F_{J}$ by varying $\Phi$.  For instance $F_{+-}<0$ for
$E,\Phi\approx0$. For $\Phi\approx 0.4\Phi_0$, where $F_{N}$ is
super-Poissonian and $F_{J}$ has a minimum, $F_{+-}>0$.

We conclude by presenting an interpretation of the influence of the
time scales on the correlations.  For this purpose, we have done a
Monte-Carlo simulation~\cite{BKL}. The time evolution of the system
making transitions with fixed conditional probability is followed. A
transition $J\to J'$ occurs with probability $\gamma_{J\to
J'}/\sum_{J''=J\pm 1}\gamma_{J\to J''}$ ($-2\leq J,J',J''\leq 2$,
$|J-J'|=1$). We consider $A\gg A_{\rm c}$ and $\Phi=0$.

\begin{figure}[htpb]
  \setlength{\unitlength}{1cm}
  \includegraphics[clip,width=7.5cm,height=3.0cm]{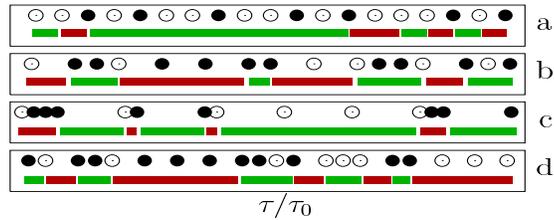}
  \vskip-0.3cm
\caption[]{Output sequences of a Monte-Carlo simulation
  for tunneling events at junction $2$, with 
  $eV=3 E_{J}$, $N_{\rm g}=0.48$, $k_{\rm B}T=0.02E_{\rm
    J}$ and $A=20$.  Black (white) dots denote  a clockwise (anticlockwise)
  entering electron. Colored bars are the  sequence of
  the orbital value $|J|$, green: oscillations $0\to\pm 1\to
  0$; red: oscillations $\pm 2\to\pm 1\to \pm 2$; (a) $g=g_{\ell}=1$; (b)
  $g=0.7$, $g_{\ell}=1$; (c) $g=0.7$, $g_{\ell}=0.5$; (d) $g=0.7$,
  $g_{\ell}=1.2$.}
\label{fig5}
\end{figure}

Figure~\ref{fig5} shows results of the simulation of tunneling events
(black and white dots) at junction $2$. Colored bars describe the
sequences of $J$ values: green bars denote $J=0$, and red bars the
excited state with $J=2$. We denote these two sequences as $S_0$ and
$S_2$. The average time interval between tunneling 
is $\tau_0$ in $S_0$ and $\tau_2$ in $S_2$ with corresponding average
number of transitions $n_0=(1+r_1)/r_1$ and $n_2=(1+r_1)$
respectively.  Only without interactions
(Fig.~\ref{fig5}a) the tunneling events are uniformly distributed
$(\tau_0=\tau_2)$. The super-Poissonian character of $F_{N}$ is due to
inhomogeneous distribution of the time scales $(\tau_0\neq \tau_2)$,
with bunches of events separated by longer times. The bunching
tendency can be present inside $S_0$ (Fig.~\ref{fig5}b, d) or inside
$S_2$ (Fig.~\ref{fig5}c). In all cases $F_{N}>1$ independent of
whether $S_0$ or $S_2$ are responsible for the bunching.  However,
bunching alone is not sufficient for $F_{N}>1$; if $A<A_{\rm c}$
bunching might still occur but the interplay of the two barriers
gives $F_{N}<1$. The interpretation of $F_{J}$ is distinctly
different. Here, the time scales $\tau_{0,2}$ do not play any role.
The important parameter is the number of event $n_2$ in which the ring
is in the excited state $J=2$. The condition $F_{J}>1$ is fulfilled
only for $n_2>2$, {\em independent of bunching}
(Fig.~\ref{fig5}d).

The simulations allow also an estimate of the correlations times for
the charge current, $T_N$, and magnetization current, $T_J$, which
determine the relaxation dynamics for long times. This, at first
glance, is only a qualitative description, which however, can
be confirmed by evaluation of the time-dependent
solutions of the master equation. The sequences $S_{0}$ and $S_{2}$
have average duration times $T_{0}=n_{0}\tau_{0}$ and
$T_{2}=n_{2}\tau_{2}$.  From Fig.~\ref{fig5}b-d, one concludes that
for the relaxation time of the angular momentum only the $S_2$
sequences contribute (red). Here $|J|=2$ before a transition to $J=0$,
such that $T_J=T_2$.  On the other hand, the charge correlation time
is determined by the two sequences and is dominated by smallest of
$T_0$ and $T_2$ since $T_{\rm N}^{-1}\equiv T_{0}^{-1}+T_{2}^{-1}$.

In summary, we have investigated zero frequency shot noise of a 1D
quantum ring attached to semi-infinite leads in the sequential
tunneling regime.  In addition to the usual charge current noise we
have identified a magnetization current shot noise which characterizes
the fluctuations in the angular current in the ring and has distinctly
different properties.  For the charge current noise a Fano factor
larger than 1 indicates bunching and super-Poissonian statistics. 
This is closely related to the presence of
interaction between the electrons in the ring and in the leads. For
the magnetization current noise it is very difficult to achieve a
Fano factor larger than one. It is independent of whether or not
bunching occurs and is  closely related to the occupation dynamics
of angular states with $J>1$.  In our example, a magnetization noise
Fano factor larger than 1 can only be reached when population
inversion is achieved and this is done by attractive interactions in
the leads. In order to verify the predicted effect experimentally, one
needs to separate from the total noise the magnetization current
contribution.  This might be done by measuring  the magnetization
\cite{grundler} of a single wall nanotube ring~\cite{shea} or
semiconductor ring~\cite{haug1} attached to leads.

We thank Bernhard Kramer for useful discussions 
and acknowledge financial support from  EU-RTN HPRN CT 2000-00144, PRIN-2002 and Firb.


\end{document}